\documentclass[11pt,a4paper]{article}
\usepackage{jcappub}
\usepackage{euscript,epsfig,amsmath,amssymb}
\usepackage{amsfonts,latexsym}
\usepackage{tabu}
\usepackage{array}
\usepackage{arydshln}
\usepackage{enumerate}

\usepackage{hyperref}
\usepackage{url}

 \newcolumntype{L}[1]{>{\raggedright\let\newline\\\arraybackslash\hspace{0pt}}m{#1}}
\newcolumntype{C}[1]{>{\centering\let\newline\\\arraybackslash\hspace{0pt}}m{#1}}
\newcolumntype{R}[1]{>{\raggedleft\let\newline\\\arraybackslash\hspace{0pt}}m{#1}}

\newcommand{\be}[1]{\begin{equation}\label{#1}}
\newcommand{\ee}{\end{equation}}
\newcommand{\ba}[1]{\begin{eqnarray}\label{#1}}
\newcommand{\ea}{\end{eqnarray}}
\newcommand{\rf}[1]{(\ref{#1})}
\newcommand{\nn}{\nonumber}

\newcommand{\sgn}{\textrm{sgn}}
\newcommand{\epsch}{\bar\varepsilon_{\mathrm{Ch}}}
\newcommand{\epschnow}{\bar\varepsilon_{\mathrm{Ch,0}}}

\extrarowsep=_6pt^8pt

\begin{document}

\title{Scalar perturbations in the late Universe: viability of the Chaplygin gas models}

\author{Mariam Bouhmadi--L\'opez$^{1,2,3,4}$,}
\author{Maxim Brilenkov$^{5}$,}
\author{Ruslan Brilenkov$^{5}$,}
\author{Jo\~ao Morais$^{3}$,}
\author{and Alexander Zhuk$^{6}$,}

\affiliation{
${}^1$ Departamento de F\'isica, Universidade da Beira Interior, 6200 Covilh\~a, Portugal \\
${}^2$ Centro de Matem\'atica e Aplica\c{c}\~oes da Universidade da Beira Interior (CMA-UBI), 6200 Covilh\~a, Portugal\\
${}^3$ Department of Theoretical Physics, University of the Basque Country
UPV/EHU, P.O. Box 644, 48080 Bilbao, Spain\\
${}^4$IKERBASQUE, Basque Foundation for Science, 48011, Bilbao, Spain\\
$^{5}$Department of Theoretical Physics, Odessa National University,\\ Dvoryanskaya st. 2, Odessa 65082, Ukraine\\
$^{6}$Astronomical Observatory, Odessa National University,\\ Dvoryanskaya st. 2, Odessa 65082, Ukraine\\
}

\emailAdd{mbl@ubi.pt (On leave of absence from UPV/EHU and IKERBASQUE)}
\emailAdd{maxim.brilenkov@gmail.com}
\emailAdd{ruslan.brilenkov@gmail.com}
\emailAdd{jviegas001@ikasle.ehu.eus}
\emailAdd{ai.zhuk2@gmail.com}

\abstract{
We study the late-time evolution of the Universe where dark energy (DE) is parametrised by a modified generalised Chaplygin gas (mGCG) on top of cold dark matter (CDM). We also take into account the radiation content of the Universe. In this context, the late stage of the evolution of the universe refers to the epoch where CDM is already clustered into inhomogeneously distributed discrete structures (galaxies, groups and clusters of galaxies). Under these conditions, the mechanical approach is an adequate tool to study the Universe deep inside the cell of uniformity. To be more accurate, we study scalar perturbations of the Friedmann-Lema\^itre-Robertson-Walker metric due to inhomogeneities of CDM as well as fluctuations of radiation and mGCG, the later driving the late-time acceleration of the universe. Our analysis applies as well to the case where mGCG plays the role of DM and DE. We select the sets of parameters of the mGCG that are compatible with the mechanical approach. These sets define prospective mGCG models. By comparing the selected sets of models with some of the latest observational data results, we conclude that the mGCG is in tight agreement with those observations particularly for a mGCG playing the role of DE and DM.
%
}

\maketitle

\flushbottom

\section{Introduction}

\

Modern observational data indicates that the universe started recently to accelerate. However, a satisfactory explanation for this acceleration has not been found so far. According to observations, something named DE constitutes about 69\%
of the total value of the energy density in the Universe \cite{Planck2015}. The nature of this constituent is a challenge for modern cosmology and theoretical physics. A number of scenarios has been proposed to explain the problem of DE, which can be divided into two groups. On the one hand, the recent speed up of the Universe could be explained through consistent modifications of gravity
\cite{Nojiri:2006ri,2012PhR...513....1C,Capozziello:2011et,DeFelice:2010aj}. For example, the nonlinear $f(R)$ theories are bright representatives of such modified theories (c.f., for example, Ref.~\cite{Morais:2015ooa} for a very recent account on those models). On the other hand, the Universe may be filled with a substance which causes the accelerated expansion \cite{Copeland:2006wr,Weinberg:1988cp,Frieman:2008sn,Padmanabhan:2002ji,Peebles:2002gy}. For example, this substance could be a quintessence field or a barotropic perfect fluid. The Chaplygin gas is one of the most popular representatives of those fluids \cite{Kamenshchik:2001cp,Bilic:2001cg,Bento:2002ps}.

Indeed, the Chaplygin gas and its generalised version could play the roles of CDM and DE \cite{Kamenshchik:2001cp,Bilic:2001cg,Bento:2002ps}. Despite this appealing fact, it has been equally shown that the generalised Chaplygin gas (GCG), unless a stringent limit is imposed on the parameter of the unified model, instabilities and oscillations on the matter power spectrum can arise from the adiabatic pressure perturbations of the GCG (c.f. for example \cite{Wang:2013qy}). Those instabilities can be avoided by considering the GCG unified model as a vacuum-CDM interacting scenario where the perturbations are non-adiabatic, being what was named the ``geodesic'' model in \cite{Wang:2013qy} a clear representative. Other approaches to overcome the GCG instabilities as a unified model can be found in \cite{Carneiro:2014jza,Borges:2013bya}. On the other hand, the GCG could simply play the role of DE and even avoid the Big Rip singularity as was shown in Ref.~\cite{BouhmadiLopez:2004me}. Furthermore, the GCG can describe some primitive epochs of the Universe \cite{Bouhmadi;fraz1,Bouhmadi;chen,Bouhmadi;mor1} and alleviate the observed low quadruple of the CMB \cite{Bouhmadi;chen2}. In this work, we will consider the most general type of Chaplygin gas, named the modified generalised Chaplygin gas (mGCG) \cite{Benaoum:2002zs}, as the substract causing the recent speed up of the Universe and in principle unrelated to the CDM component in the Universe.

Obviously, to explain the late time acceleration of the Universe, the Chaplygin gas model or any of its different versions should be valid for a certain period of time, starting from some early time, roughly since red-shift $z\sim O(10) $, and continuing for some future period with respect to the present time.
In our present work, we consider the case when the mGCG induces no future singularities (related to a divergence of the energy density or pressure of the DE component) in the Universe \cite{Nojiri:2005sx,BouhmadiLopez:2006fu,BouhmadiLopez:2007qb}. Therefore, the mGCG is valid for arbitrarily big values of the scale factor, $a$. This means that we consider models where the scale factor of the Universe does not reach a finite maximum allowed value; e.g. like in a Big Freeze singularity \cite{BouhmadiLopez:2006fu}
\footnote{\label{footnote1}{In the mechanical approach, the functions describing, e.g., the energy density and the pressure for the different matter sources as well as their fluctuations are expanded in powers of the inverse of the scale factor, $1/a$. This procedure is mathematically well defined if the scale factor can go to infinity; i.e. $1/a=0$ is the point of expansion. On that case, we know for sure that the terms proportional to $(1/a)^n$ can be disregarded asymptotically as compared with $(1/a)^m$, if $m<n$. If the scale factor of the Universe has a maximum size in its future expansion, the validity of the mechanical procedure is no longer guaranteed and a separate and additional investigation is required. See also footnote \ref{foot2} and the paragraph related to that footnote.}}. We consider the Universe at the late stage of its evolution when the galaxies and the group of galaxies are already formed. At this stage, and inside the cell of uniformity, the Universe is highly inhomogeneous and the mechanical approach \cite{EZcosm1,EKZ2,EZcosm2} is an adequate approximation for its description. The theory of scalar perturbations within the mechanical approach was proposed in \cite{EZcosm1,EZcosm2}. One of the main advantages of this approach consists in the fact that it gives the possibility to investigate analytically different cosmological models from the point of view of their compatibility with the theory of scalar perturbations (see e.g. \cite{BUZ1,Laslo2,ENZ1}).

The main idea of such investigations is the following. As we noted above, the considered cosmological models are valid for a certain period of time. It is clear that the scalar perturbation equations must be consistent during all that period. Concerning the mechanical approach, its accuracy becomes better and better with the growth of the scale factor $a$. If the scale factor of the models is not limited from above, then we can investigate the equations of the scalar perturbations for arbitrarily big values of $a$. This allows us to simplify considerably our analysis and look for a set of parameters (if such set exists) compatible with the scalar perturbation equations.
The models with these sets of parameters are expected to be viable as they are consistent with the theory of scalar perturbations for big values of $a$ and at the same time can provide a period of acceleration at present. Then, the models selected as viable should be tested (e.g. for different ranges of $a$ or redshift) against other observational data.
If the parameters do not belong to these sets, the corresponding models are unsuitable models in the future. This means that
such parametrizations can be used only for a limited period of time, starting e.g. from the time of the last scattering surface till present. In our paper we find a number of combinations of the parameters for the mGCG models which are consistent with the equations of the scalar perturbations in the limit of big scale factors (i.e. in the late Universe). These sets define prospective Chaplygin gas models.

The paper is structured as follows. In Sec.~\ref{mGCG} we review the mGCG cosmology. In Sec.~\ref{Pert}, we consider the theory of the scalar perturbation for the mGCG within the mechanical approach and find the sets of parameters for which the equations of the scalar perturbations are consistent. The main results are summarised and discussed in the concluding Sec.~\ref{Conc}.


\section{Reviewing the mGCG}
\label{mGCG}

\setcounter{equation}{0}

\

The barotropic perfect fluid called the mGCG is characterized by the equation of state (EoS) \cite{Benaoum:2002zs}
\be{2.1}
	\bar p_{\mathrm{Ch}}
		= \beta \epsch
		- (1+\beta)\frac{A}{\bar\varepsilon^{\alpha}_{\mathrm{Ch}}}\, ,
\ee
where $\beta$, $A$ and $\alpha$ are the constant parameters of the model. The mechanical approach for the particular cases $\alpha=-1$ as well as $A=0$ were considered in our previous papers \cite{BUZ1,CPL} and we exclude them here.

Within the parametrization \rf{2.1}, the case $\beta=-1$ is an exceptional one, corresponding to the EoS of a cosmological constant. Obviously, if $\beta \neq -1$ the equation of state \rf{2.1} is equivalent to the following one:
\be{2.1a}
	\bar p_{\mathrm{Ch}}
		= \beta \epsch
		- \frac{\tilde A}{\bar\varepsilon^{\alpha}_{\mathrm{Ch}}}\, ,
\ee
where $\tilde A=(1+\beta)A$. If we ignore this equality, we can consider \rf{2.1a} as an independent parametrization (from the EoS \rf{2.1}) where the case $\beta =-1$ is not reduced to the cosmological constant. This particular case will be studied in subsection \ref{LSBR} below.

Let us investigate first the case $\beta\neq -1$ and $\alpha\neq -1$. Then, the conservation equation results in the following dependence of the energy density on the scale factor $a$:
\be{2.2}
	\epsch
		=\epschnow\left[
			A_s
			+\left(1-A_s\right)\left(\frac{a_0}{a}\right)^{3\left(1+\beta\right)\left(1+\alpha\right)}
		\right]^{\frac{1}{1+\alpha}}\, ,
\ee
where $a_0$ and $\epschnow$ are the current values of the scale factor and the energy density of the mGCG, and $A_s\equiv A/\epschnow^{1+\alpha}$. Here, we exclude the case $A_s=0$, as it corresponds to $A=0$, and the case $A_s=1$ because it reduces to the trivial cosmological constant case.

 We next analyse the behaviour of the mGCG depending on the values of its parameters $A_s$, $\alpha$, and $\beta$. If we define the quantities $a_*$ and $\xi$ as
\ba{new1}
	a_*\equiv a_0 \left|\frac{1-A_s}{A_s}\right|^{\frac{1}{3(1+\beta)(1+\alpha)}},
	~~~~~~~~\textrm{and}~~~~~~~~
	\xi \equiv (1+\beta)(1+\alpha)\, ,
\ea
 we can rewrite Eq.~\eqref{2.2} as
\ba{new2}
	\epsch
		= \epschnow\left|A_s\right|^{\frac{1}{1+\alpha}}
		\left[\sgn\left(A_s\right) + \sgn\left(1-A_s\right)\left(\frac{a_*}{a}\right)^{3\xi}\right]^{\frac{1}{1+\alpha}}\, .
\ea
Here, $\sgn(x)$ is the sign function which returns $1$ $(-1)$ if $x$ is positive (negative) and $0$ if $x=0$. Notice that since $A_s\neq0,1$, $a_*$ is a finite positive quantity, and that $\xi\neq0$ as long as $\alpha\neq-1$ and $\beta\neq-1$.

From Eq.~\eqref{new2} we find that in order for the energy density to be well defined, the right-hand-side (rhs) of such an equation must be positive. This is ensured by setting $(1+\alpha)^{-1}=2n$, where $n$ is an integer, or, for a general $\alpha$, by requiring that the quantity inside the square brackets be positive,
\ba{new3}
	\sgn(A_s) + \sgn(1-A_s)\left(\frac{a_*}{a}\right)^{3\xi} >0.
\ea
When we analyse this inequality, we find the following restrictions for the value of the scale factor, summarised in Table~\ref{mGCG_restrictions},
\begin{enumerate}[(i)]
\item For $A_s<0$ the energy density of the mGCG is only defined either to the left or right of the value $a_*$, depending on the sign of $\xi$.

\item For $0<A_s<1$ the energy density of the mGCG is defined for all values of the scale factor.

\item For $A_s>1$ the energy density of the mGCG is only well defined to the left or right of $a_*$, depending on the sign of $\xi$.
\end{enumerate}


\begin{table}[t]
\centering
\begin{tabu}{ C{.22\textwidth} : C{.22\textwidth} : C{.22\textwidth} : C{.22\textwidth} }
				& $A_s<0$			& $0<A_s<1$		& $1<A_s$ \\ \hline
	$\xi<0$		& $a_*<a<+\infty$		& $0<a<+\infty$ 	& $0<a<a_*$ \\
	$0<\xi$		& $0<a<a_*$			& $0<a<+\infty$	& $a_*<a<+\infty$ \\ \hline
\end{tabu}
\caption{\label{mGCG_restrictions}Intervals of the value of the scale factor, $a$, for which the inequality \eqref{new3} is satisfied.}
\end{table}

\begin{table}[t]
\centering
	\begin{tabu}{ C{.47\textwidth} : C{.47\textwidth} }
	\( \xi<0 \)		& \( 0<\xi\) \\ \hline
	\( a_*<a<+\infty \) 	& \( 0<a<a_* \) \\
	\( a_*\ll a \Rightarrow \frac{\epsch(a)}{\epschnow} \approx \left|A_s\right|^{\frac{1}{1+\alpha}}\left(\frac{a_*}{a}\right)^{\frac{3\xi}{1+\alpha}} \)
		& \( a\ll a_* \Rightarrow \frac{\epsch(a)}{\epschnow} \approx \left|A_s\right|^{\frac{1}{1+\alpha}}\left(\frac{a_*}{a}\right)^{\frac{3\xi}{1+\alpha}} \)\\
	\(\alpha<-1 \Rightarrow \lim\limits_{a\to a_*}\epsch(a) = +\infty\)
		& \( \alpha<-1 \Rightarrow \lim\limits_{a\to a_*}\epsch(a) = +\infty\) \\
	\( -1 < \alpha \Rightarrow \lim\limits_{a\to a_*}\epsch(a) = 0\)
		& \( -1 < \alpha \Rightarrow \lim\limits_{a\to a_*}\epsch(a) = 0 \) \\ \hline
\end{tabu}
\caption{\label{behav_l0}Behaviour of the mGCG for $A_s<0$.}
\end{table}
\begin{table}
\centering
	\begin{tabu}{ C{.47\textwidth} : C{.47\textwidth} }
	\( \xi<0 \)		& \( 0<\xi\) \\ \hline
	\( 0<a<+\infty \) 	& \( 0<a<+\infty \) \\
	\( a\ll a_* \Rightarrow \frac{\epsch(a)}{\epschnow} \approx \left|A_s\right|^{\frac{1}{1+\alpha}} \)
		& \( a\ll a_* \Rightarrow \frac{\epsch(a)}{\epschnow} \approx \left|A_s\right|^{\frac{1}{1+\alpha}}\left(\frac{a_*}{a}\right)^{\frac{3\xi}{1+\alpha}} \)\\
	\( a_*\ll a \Rightarrow \frac{\epsch(a)}{\epschnow} \approx \left|A_s\right|^{\frac{1}{1+\alpha}}\left(\frac{a_*}{a}\right)^{\frac{3\xi}{1+\alpha}} \)
		& \( a_*\ll a \Rightarrow \frac{\epsch(a)}{\epschnow} \approx \left|A_s\right|^{\frac{1}{1+\alpha}}\)\\
	\(\alpha\neq-1 \Rightarrow \frac{\epsch(a_*)}{\epschnow} = \left|2A_s\right|^{\frac{1}{1+\alpha}}\)
		& \(\alpha\neq-1 \Rightarrow \frac{\epsch(a_*)}{\epschnow} = \left|2A_s\right|^{\frac{1}{1+\alpha}}\) \\ \hline
\end{tabu}
\caption{\label{behav_01}Behaviour of the mGCG for $0<A_s<1$.}
\end{table}
\begin{table}[t]
\centering
	\begin{tabu}{ C{.47\textwidth} : C{.47\textwidth} }
	\( \xi<0 \)		& \( 0<\xi\) \\ \hline
	\( 0<a<a_* \) 	& \( a_*<a<\infty \) \\
	\( a\ll a_* \Rightarrow \frac{\epsch(a)}{\epschnow} \approx \left|A_s\right|^{\frac{1}{1+\alpha}}\left(\frac{a_*}{a}\right)^{\frac{3\xi}{1+\alpha}} \)
		& \( a_*\ll a \Rightarrow \frac{\epsch(a)}{\epschnow} \approx \left|A_s\right|^{\frac{1}{1+\alpha}}\left(\frac{a_*}{a}\right)^{\frac{3\xi}{1+\alpha}} \)\\
	\(\alpha<-1 \Rightarrow \lim\limits_{a\to a_*}\epsch(a) = +\infty\)
		& \( \alpha<-1 \Rightarrow \lim\limits_{a\to a_*}\epsch(a) = +\infty\) \\
	\( -1 < \alpha \Rightarrow \lim\limits_{a\to a_*}\epsch(a) = 0\)
		& \( -1 < \alpha \Rightarrow \lim\limits_{a\to a_*}\epsch(a) = 0 \) \\ \hline
\end{tabu}
\caption{\label{behav_g1}Behaviour of the mGCG for $1<A_s$.}
\end{table}

Having obtained the regions where the energy density of the mGCG is well defined, we can study its behaviour near the limits of those regions. Notice that when $A_s\not\in (0,1)$ and $1+\alpha<0$, the mGCG has a singularity at $a=a_*$. For $A_s\in(0,1)$ the mGCG interpolates smoothly between a cosmological constant and a perfect fluid with a constant parameter of EoS, with $a_*$ marking the transition point between these two behaviours. Notice that the effective cosmological constant behaviour of the mGCG happens in the future (past) whenever $\xi>0$ ($\xi<0$) (c.f. Eq. ~\eqref{new2}). The results obtained are summarized in Tables~\ref{behav_l0}, \ref{behav_01}, and \ref{behav_g1}. We will restrict our analysis to those cases where the mGCG energy density is well defined as $a\rightarrow+\infty$. For these cases, we obtain late time acceleration whenever $\xi>0$, where, as stated above, the mGCG mimics a cosmological constant; or when $\xi<0$ and $\beta<-1/3$, where the mGCG behaves as a perfect fluid with constant parameter of EoS $\beta$. Notice that in the latter we obtain phantom behaviour at late time whenever $\beta<-1$ (because the asymptotic effective parameter of the EoS is proportional to $\beta$).


 \section{Scalar perturbations: suitable and non-suitable cases}
\label{Pert}

\

The theory of the scalar perturbations for the late Universe filled with a barotropic perfect fluid with an arbitrary EoS $\bar p_{\mathrm{X}}= \bar p_{\mathrm{X}}( \bar\varepsilon_{\mathrm{X}})$ was recently considered in our paper \cite{CPL}, where Eqs.~(2.17) and~(2.18) of that paper are the master equations{\footnote{\label{coupled} Fluctuations of an additional perfect fluid (which can be responsible for the late-time acceleration of the Universe) arise due to inhomogeneities of dust-like matter (e.g., galaxies). These fluctuations also form their own inhomogeneities. In the mechanical approach, it is supposed that the velocities of displacement of such inhomogeneities is of the order of the peculiar velocities of inhomogeneities of dust-like matter, i.e. they are non-relativistic. In some sense, these two types of inhomogeneities are "coupled" to each other \cite{RGS}. As we can see from the master equation below, this results in a change of the gravitational potential. The flattening of rotation curves can be one of the consequences of such a change \cite{Laslo2}.}. In the case of a mGCG, these equations read
\ba{2.3}
	&{}&-\left(\bar{\varepsilon}_{\mathrm{Ch}}+\bar{p}_{\mathrm{Ch}}\right)\frac{\varphi }{c^2 a}
		=\delta p_{\mathrm{Ch}}+ \frac{1}{3}\delta\varepsilon_{\mathrm{rad2}}\, , \\
		&{}& \label{2.4} \triangle\varphi+3\mathcal{K}\varphi
		-\frac{\kappa c^4}{2}\delta\rho_{\mathrm{ c}}
		=\frac{\kappa c^2 a^3}{2}\delta\varepsilon_{\mathrm{Ch}}+\frac{\kappa
c^2 a^3}{2}\delta\varepsilon_{\mathrm{rad2}}\, ,
\ea
where $\kappa\equiv 8\pi G_N/c^4$ ($c$ is the speed of light and $G_N$ is the Newton's gravitational constant),
$\mathcal K=-1,0,+1$ for open, flat and closed Universes, respectively, and the Laplace operator $\triangle$ is defined via the conformal spatial metrics.
Here, $\varphi$ is the comoving gravitational potential and $\delta\rho_{\mathrm{\, c}}$ is the difference between the real and average comoving rest mass densities{\footnote{\label{dimen} The comoving rest mass density and the scale factor have dimensions of mass and length, respectively \cite{EZcosm1}. It is worth noting also that the combinations $\varphi/(c^2 a)$ and $\kappa a^2 \varepsilon$ are dimensionless.}}: $\delta\rho_{\mathrm{ c}} = \rho_{\mathrm{\, c}}-\bar\rho_{\mathrm{\, c}}$ of inhomogeneities in the form of galaxies and group of galaxies which may or may not include CDM, depending on whether the mGCG plays the role of a unified model or just the role of DE (see details in \cite{EZcosm1,EZcosm2}). The analytical results of our work do not depend on which of the two background models we are considering. However, the comparison to the observations that we do on the Conclusions applies only to the unified model. Both $\varphi$ and $\delta\rho_{\mathrm{\, c}}$ do not depend explicitly on time. $\delta\varepsilon_{\mathrm{Ch}}$ and $\delta p_{\mathrm{Ch}}$ describe fluctuations of the mGCG energy density and pressure, respectively. We also take into account the presence of radiation in the Universe. This means that to derive Eqs.~\rf{2.3} and \rf{2.4}, we consider terms up to $O(1/a^4)$ inclusive and neglect those of the order $o(1/a^{4})$. This is the accuracy of our study.
We also consider the possibility that the inhomogeneities and the mGCG fluctuations can contribute to the fluctuations of radiation. Therefore, we introduce $\delta\varepsilon_{\mathrm{rad1}}$ and $\delta\varepsilon_{\mathrm{rad2}}$ which are the fluctuations of radiation associated with the matter inhomogeneities and the mGCG fluctuations, respectively. Of course, experiments measure the total fluctuation of radiation. Nevertheless, theoretically we can consider these contributions separately. In our paper \cite{EZcosm2}, we have shown that $\delta\varepsilon_{\mathrm{rad1}}=-3\bar{\rho}_{\mathrm{c}}\,\varphi/a^4$. This relation is already taken into account in Eqs.~\rf{2.3} and \rf{2.4}.
With respect to the mGCG, the contribution to the total $\delta\varepsilon_{\mathrm{rad}}$ may take place directly via non-zero $\delta\varepsilon_{\mathrm{rad2}}$ in Eqs.~\rf{2.3} and \rf{2.4}. There is also the possibility that the mGCG contributes to the radiation fluctuations even if $\delta\varepsilon_{\mathrm{rad2}}\equiv 0$. This can be easily seen from Eq.~\rf{2.4} where the comoving gravitational potential depends on the presence of $\delta\varepsilon_{\mathrm{Ch}}$. On the other hand, as we have mentioned above, $\varphi$ defines $\delta\varepsilon_{\mathrm{rad1}}$. Given these two possible scenarios, we include
$\delta\varepsilon_{\mathrm{rad2}}$ into our equations.
For these two radiative fluctuation contributions, we have the same equations of state: $\delta p_{\mathrm{rad1}}=(1/3)\delta\varepsilon_{\mathrm{rad1}},\quad \delta
p_{\mathrm{rad2}}=(1/3)\delta\varepsilon_{\mathrm{rad2}}\, .$ Therefore, from the matter conservation equation we get
$\delta\varepsilon_{\mathrm{rad2}} \sim 1/a^4$ (see the appendix in Ref.~\cite{CPL}).

From the mGCG EoS \rf{2.1} we get
\be{2.6}
	\delta p_{\mathrm{Ch}}
		=\frac{
			A_s(\xi-1) + (1-A_s)\beta (\frac{a_0}{a})^{3\xi}
		}
		{
			A_s + (1-A_s) (\frac{a_0}{a})^{3\xi}
		}
		\delta\varepsilon_{\mathrm{Ch}}
	\, .
\ee
Then, taking into account Eqs.~\rf{2.1} and \rf{2.2}, we get from Eq.~\rf{2.3} the expression for $\delta\varepsilon_{\mathrm{Ch}}$:
\ba{2.8}
	\delta\varepsilon_{\mathrm{Ch}}
		&=& -
		\left[
			A_s
			+ (1-A_s)\left(\frac{a_0}{a}\right)^{3\xi}
		\right]^{\frac{1}{1+\alpha}}
		\frac{(1-A_s)(1+\beta)(\frac{a_0}{a})^{3\xi}}{A_s(\xi-1) + (1-A_s)\beta (\frac{a_0}{a})^{3\xi}}
		\frac{\epschnow}{c^2 a}\varphi\nn \\
		&{ }&-\frac{1}{3}
		\frac{
			A_s
			+ (1-A_s)\left(\frac{a_0}{a}\right)^{3\xi}
		}
		{
			A_s(\xi-1) + (1-A_s)\beta\left(\frac{a_0}{a}\right)^{3\xi}
		}
		\delta\varepsilon_{\mathrm{rad2}}
		\, .
\ea

Now, we can investigate the consistency of Eq.~\rf{2.4} with $\delta\varepsilon_{\mathrm{Ch}}$ given in Eq.~\rf{2.8} for different values of $\xi$. Before tackling this analysis, we want to make the following remark. As we mentioned above, the system of Eqs.~\rf{2.3} and \rf{2.4} was derived for the late Universe, i.e. for rather big values of the scale factor $a$, starting e.g. from the present moment. In these equations we neglect terms $o(1/a^4)$ (see footnote 2 in \cite{CPL}). Obviously, the larger the value of $a$, the more accurate our mechanical approach is.
That is, it should work perfectly in the future Universe. On the other hand, Eq.~\rf{2.8} and consequently \rf{2.4} are considerably simplified in the approximation of big values of the scale factor, $a$, as we are neglecting terms of the order $o(1/a^{4})$. Following this prescription, we first expand the expression \rf{2.8} in powers of $1/a$, neglecting terms $o(1/a^4)$. Then we substitute this expression into Eq.~\rf{2.4} and analyse the consistency of the equation obtained. The equation is considered consistent if its rhs does not depend on time up to terms $O(1/a)$ inclusive (bearing in mind that the rhs of this equation has been multiplied by $a^3$). It is clear that the result depends on the values of the parameters of the model. If we can define the sets of parameters which make this equation consistent, then we could claim that such models are possible candidate for prospective Chaplygin gas realisations{\footnote{\label{foot2}{We perform this analysis for big values of the scale factor. It is clear that the candidates found should be investigated further for smaller values of $a$, e.g. at the present moment where the discrete cosmology approach is already valid. We need to compare terms (of the order $O(1/a^4)$) which were left in Eq.~\rf{2.8} with the ones we dropped (i.e. $o(1/a^4)$). Of course, we can do it if we know the observable values of the parameters of the model.
This is a general remark. However, if a perfect fluid such as the Chaplygin gas is responsible for the present day acceleration of the Universe, it is clear that terms $O(1/a^5)$ should not play an important role now. According to the observations, they must contribute less than radiation, i.e. terms of the order of $O(1/a^4)$.}}}. As we can see from Eq.~\rf{2.8}, the powers of the ratio $(1/a)$ are mainly defined by the parameter $\xi$. Therefore, it makes sense to investigate this equation separately for different values of $\xi$.

\

\subsection{$\xi>1$}
\label{gr1_xi}
\

In this case, for large values of the scale factor, $a$, Eq.~\rf{2.8} can be approximated as
\be{2.10}
	\delta \varepsilon_{\mathrm{Ch}}=-\frac{1}{3}\frac{1}{\xi -1}\delta\varepsilon_{\mathrm{rad2}}+ o(1/a^4)
	\, .
\ee
Therefore, Eq.~\rf{2.4} reads
\be{2.11}
	\triangle\varphi+3\mathcal{K}\varphi-\frac{\kappa c^4}{2}\delta\rho_{\mathrm{c}}
	=\frac{\kappa c^2 a^3}{2}
	\left[
		1
		-\frac{1}{3(\xi -1)}
	\right]
	\delta\varepsilon_{\mathrm{rad2}}
	\, ,
\ee
where we dropped out the terms $o(1/a)$ (bearing in mind that the rhs of \rf{2.4} has been
multiplied by $a^3$). The left hand side (lhs) of this equation does not depend on time{\footnote{ We would like to remind that the quantities $\varphi$ and $\delta\rho_{\mathrm{ c}}$ are the comoving ones \cite{EZcosm1}. Therefore, within the adopted accuracy when both non-relativistic and weak field limits are applied, they do not depend explicitly on time \cite{EZcosm2}.}}. Therefore, the rhs should also not depend on time.
Taking into account that $\delta\varepsilon_{\mathrm{rad2}} \sim 1/a^4$, we can achieve this in two ways. Either $\delta\varepsilon_{\mathrm{rad2}}=0$, which implies that $\delta \varepsilon_{\mathrm{Ch}}=0$ (up to the adopted accuracy) and then the mGCG does not cluster, which is physically less reasonable, or
\be{2.12}
	\xi=\frac43 \quad \Longrightarrow \quad \alpha=\frac{1-3\beta}{3(1+\beta)}
	\, ,
\ee
and the rhs of \rf{2.11} is equal to zero.
In this particular case, $\delta\varepsilon_{\mathrm{rad2}}$ can be non-zero, which results in non-zero values of $\delta \varepsilon_{\mathrm{Ch}}$, since $\delta \varepsilon_{\mathrm{Ch}} \sim \delta\varepsilon_{\mathrm{rad2}}$ (see Eq.~\rf{2.10}). That means the mGCG is clustered.

For example, in the particular case $\beta=0 \, \Rightarrow \, \alpha = 1/3$ we get
\ba{}
	\bar p_{\mathrm{Ch}}=-\epschnow A_s\left(\frac{\epschnow}{\epsch}\right)^{\frac{1}{3}}
	\, ,
	 ~~~~&\textrm{and}&~~~~
	\epsch = \epschnow\left[A_s + (1-A_s)\left(\frac{a_0}{a}\right)^{4}\right]^{\frac{3}{4}}
 \,.
 \ea
This fluid behaves as dust for small values of $a$: $\epsch\sim 1/a^3$ (if $0<A_s <1$); and as a cosmological constant for large values of $a$: $\epsch\sim const$, in full accordance with Table~\ref{behav_01} (right column). Another interesting example corresponds to the choice $\alpha=0 \, \Rightarrow \, \beta = 1/3$. Here,
\ba{}
	\bar p_{\mathrm{Ch}}=\frac{\epsch-4A_s\epschnow}{3}
	\, ,
	~~~~&\textrm{and}&~~~~
	\epsch = \epschnow\left[A_s + (1-A_s)\left(\frac{a_0}{a}\right)^{4}\right]
	\, .
\ea
This type of model behaves as a mixture of radiation and a cosmological constant and it has been applied to the early Universe in our works
\cite{Bouhmadi;fraz1,Bouhmadi;mor1}.

\

\subsection{$\xi=1, \beta\neq 0$}
\label{eq1_xi}

\

The case $\xi=1$, $\beta= 0$, implies $\alpha= 0$ and describes the mixture of the cosmological constant and dust, i.e. the $\Lambda$CDM model. Both of these terms (the cosmological constant and dust in the form of inhomogeneities) have already been taken into account in the derivation of Eqs.~\rf{2.3} and \rf{2.4} (see also \cite{EZcosm1}). Therefore, we exclude this case.

If we consider the case $\beta\neq 0$ while neglecting terms of order $o(1/a^4)$ in the approximation of large $a$, we get from \rf{2.8}
\ba{2.13}
	\delta\varepsilon_{\mathrm{Ch}}
		&\approx&
		-\frac{1+\beta}{\beta}
		A_s^{\beta}
		\left[
			A_s + (1+\beta)(1-A_s)\left(\frac{a_0}{a}\right)^3	
		\right]
		\frac{\epschnow}{c^2 a}\varphi\nn \\
		&{ }&-
		\frac{1}{3\beta}
		\left[
			1
			+\frac{A_s}{1-A_s}\left(\frac{a}{a_0}\right)^3
		\right]	
		\delta\varepsilon_{\mathrm{rad2}}\, .
\ea
Then, Eq.~\rf{2.4} takes the form
\ba{2.14}
	\triangle\varphi + 3\mathcal{K}\varphi - \frac{\kappa c^4}{2}\delta\rho_{\mathrm{c}}
		&=&
		-\frac{1+\beta}{\beta}A_s^{1+\beta}
		{
		\frac{\kappa a^2 \epschnow}{2}
		}
		\varphi
		-\frac{(1+\beta)^2}{\beta} \left(1-A_s\right)A_s^\beta
		{
		\left(\frac{a_0}{a}\right)^3
		\frac{\kappa a^2 \epschnow}{2}
		}
		\varphi
		\nn \\
		&{ }&
		-\frac{1}{3\beta}\frac{A_s}{1-A_s}
		{
		\left(\frac{a_0}{a}\right)^{-3}
		\frac{\kappa c^2 a^3}{2}
		}
		\delta\varepsilon_{\mathrm{rad2}}
		+\left(1-\frac{1}{3\beta}\right)
		{
		\frac{\kappa c^2 a^3}{2}
		}
		\delta\varepsilon_{\mathrm{rad2}}
		\, .
\ea
Because $\delta\varepsilon_{\mathrm{rad2}} \sim 1/a^4$ and the comoving gravitational potential $\varphi$ does not depend on time (as well as on $a$), this equation is consistent only if the following system of equations are compatible:
\ba{2.15}
	&{}&
	-\frac{1+\beta}{\beta}A_s^{1+\beta}
	{
	\frac{\kappa a^2 \epschnow}{2}
	}\varphi
	-\frac{1}{3\beta}\frac{A_s}{1-A_s}
	{
	\left(\frac{a_0}{a}\right)^{-3}\frac{\kappa c^2 a^3}{2}
	}
	\delta\varepsilon_{\mathrm{rad2}}
	=0\, ,
	\\
	\label{2.16}
	&{}&
	-\frac{(1+\beta)^2}{\beta} \left(1-A_s\right)A_s^\beta
	{
	\left(\frac{a_0}{a}\right)^3
	\frac{\kappa a^2 \epschnow}{2}
	}\varphi
	+\left(1-\frac{1}{3\beta}\right)
	{
	\frac{\kappa c^2 a^3}{2}
	}
	\delta\varepsilon_{\mathrm{rad2}}
	=0\, .
\ea
{ It takes place e.g. if $\delta\varepsilon_{\mathrm{rad2}}=\varphi=0$. However, this is a non-physical result because the presence of inhomogeneities in the form of galaxies and group of galaxies in the Universe must result in a non-zero gravitational potential: $\varphi \neq 0$. Therefore, we exclude such solution. Then, from \rf{2.15} and \rf{2.16} we get respectively
\be{2.17}
\delta\varepsilon_{\mathrm{rad2}}=-3(1+\beta)(1-A_s)A_s^{\beta}\left(\frac{a_0}{a}\right)^3\frac{\varphi}{c^2a}\epschnow
\ee
and
\be{2.17a}
\delta\varepsilon_{\mathrm{rad2}}= \frac{3(1+\beta)^2}{3\beta-1}(1-A_s)A_s^{\beta}\left(\frac{a_0}{a}\right)^3\frac{\varphi}{c^2a}\epschnow\, .
\ee
It can be easily seen that these equations are compatible if $\beta=0$. However, this contradicts the condition $\beta \neq 0$.
Therefore, the case $\xi=1$, $\beta\neq 0$ is {\bf forbidden} within the mechanical approach.}

\

\subsection{$0<\xi<1$}
\label{0t1_xi}

\

Here, for big values of the scale factor, $a$, the mGCG fluctuation \rf{2.8} is approximated as
\be{2.18}
	\delta\varepsilon_{\mathrm{Ch}}
	\approx
	\frac{1+\beta}{1-\xi}\frac{1-A_s}{A_s}A_s^{\frac{1}{1+\alpha}}
	{ \left(\frac{a_0}{a}\right)^{3\xi}
	\frac{\epschnow}{c^2 a}}
	\varphi
	+ \frac{1}{3(1-\xi)}\delta\varepsilon_{\mathrm{rad2}}
	\, ,
\ee
and after substituting it in Eq.~\rf{2.4} we get
\ba{2.19}
	\triangle\varphi+3\mathcal{K}\varphi-\frac{\kappa c^4}{2}\delta\rho_{\mathrm{c}}&=&
	\frac{1+\beta}{1-\xi}\frac{1-A_s}{A_s}A_s^{\frac{1}{1+\alpha}}
	{ \left(\frac{a_0}{a}\right)^{3\xi}
	\frac{\kappa a^2\epschnow}{2}}\varphi
	\nn\\
	&{}&+\left(
		1
		+ \frac{1}{3(1-\xi)}
	\right)
	{\frac{\kappa c^2 a^3}{2}}
	\delta\varepsilon_{\mathrm{rad2}}
\, .
\ea
Similar to the previous cases, we need to find conditions under which the rhs of this equation is independent of time.
Because $-2<3\xi-2<1$, the first term on the rhs in Eq.~\eqref{2.19} cannot be compensated by the second one, which behaves as $\delta\varepsilon_{\mathrm{rad2}} \sim 1/a^4$ \cite{CPL}. The only way to solve this problem is to put $\delta\varepsilon_{\mathrm{rad2}}\equiv 0$ and to make the first term in Eq.~\eqref{2.19} independent on time. The latter condition takes place if
$3\xi-2=0\, \Rightarrow \, \xi =2/3$. Then, the mGCG fluctuation reads
\be{2.21}
	\delta\varepsilon_{\mathrm{Ch}}=
	3(1+\beta)\frac{1-A_s}{A_s}A_s^{\frac{1}{1+\alpha}}
	{\left(\frac{a_0}{a}\right)^{2}
	\frac{\epschnow}{c^2 a}}\varphi
	\, .
\ee
Therefore, the viable set of parameters is
\be{2.22}
	\xi = \frac23 \quad \Longrightarrow \quad \alpha= - \frac{1+3\beta}{3(1+\beta)}
	\, .
\ee
A particularly interesting case is $\beta=-1/3 \, \Rightarrow \, \alpha =0$, where
\ba{}
	\bar p_{\mathrm{Ch}}=-\frac{\epsch+2A_s\epschnow^{ }}{3}
	\, ,
	~~~~&\textrm{and}&~~~~
	\epsch=\epschnow^{ }\left[A_s+(1-A_s)\left(\frac{a_0}{a}\right)^2\right]
	\, .
\ea
It describes a mixture of a perfect fluid with the parameter of EoS $-1/3$ (e.g. frustrated network of cosmic strings) and a cosmological constant. This perfect fluid was considered in our papers~\cite{BUZ1,CPL} and is a specific case of the model we used in \cite{Bouhmadi;chen2}. It can be easily seen from Eq.~\rf{2.21} that for this particular case $\delta\varepsilon_{\mathrm{Ch}}=2(1-A_s){ a_0^2\epschnow^{ }\varphi/(c^2a^3)}$ which exactly coincides with Eq.~(2.26) in Ref.~\cite{BUZ1} after the evident substitution ${\varepsilon_0 \equiv (1-A_s)\epschnow}$.

\

\subsection{$\xi <0$}
\label{neg_xi}

\

Originally, the Chaplygin gas model was introduced into cosmology to describe the late time acceleration of the Universe.
However, in the case of negative $\xi$, the mGCG behaves as vacuum: $\bar p_{\mathrm{Ch}}=-\bar \varepsilon_{\mathrm{Ch}} \sim $const for small scale factors $a$, and as some form of matter with the energy density $\bar \varepsilon_{\mathrm{Ch}} \sim a^{-3(1+\beta)}$ for large $a$ (e.g. as a dust, radiation and frustrated network of cosmic strings or spatial curvature for $\beta=0$, $\beta=1/3$ and $\beta=-1/3$, respectively.). Therefore, such models are more suitable for the description of the early inflationary era rather than the late acceleration as was done shown in Refs.~\cite{Bouhmadi;fraz1,Bouhmadi;chen,Bouhmadi;mor1,ChInfl,Bouhmadi;chen2}. For the sake of completeness, we also investigate this case.

In the case $\xi<0$, it is convenient to introduce a new parameter $\mu \equiv -\xi >0$. Then, Eq.~\rf{2.8} reads
\ba{2.23}
	\delta\varepsilon_{\mathrm{Ch}}
		&=&
		\left[
			(1-A_s)
			+A_s\left(\frac{a_0}{a}\right)^{3\mu}
		\right]^{\frac{1}{1+\alpha}}
		\frac{(1-A_s)(1+\beta)}{A_s(\mu+1)(\frac{a_0}{a})^{3\mu} - (1-A_s)\beta}
		{\left(\frac{a_0}{a}\right)^{3(1+\beta)}
		\frac{\epschnow}{c^2 a}}\varphi\nn \\
		&{ }&+\frac{1}{3}
		\frac{
			A_s\left(\frac{a_0}{a}\right)^{3\mu}
			+ (1-A_s)
		}
		{
			A_s(\mu+1)\left(\frac{a_0}{a}\right)^{3\mu}
			- (1-A_s)\beta
		}
		\delta\varepsilon_{\mathrm{rad2}}
		\, .
\ea
It makes sense to conduct a further analysis, depending on the sign of the parameter $\beta$.

\

{\em i) $\beta>0$}

\

Obviously, for $\beta>0$ we have
\be{2.24}
	\delta\varepsilon_{\mathrm{Ch}} \approx -\frac{1}{3\beta} \delta\varepsilon_{\mathrm{rad2}} + o(1/a^4)\, .
\ee
Therefore, Eq.~\rf{2.4} is consistent for $\beta=1/3$ when its rhs is equal to zero within the accuracy of our approach. Therefore, non-zero values of $\delta\varepsilon_{\mathrm{rad2}}$ and $\delta\varepsilon_{\mathrm{Ch}}$ are allowed in this particular case.
Obviously, there is also the less physically motivated case $\beta\neq 1/3$ where $\delta\varepsilon_{\mathrm{rad2}}=\delta\varepsilon_{\mathrm{Ch}}=0$. In this case, the condition $\delta\varepsilon_{\mathrm{Ch}}=0$ means that the Chaplygin gas is not clustered.

\

{\em ii) $\beta=0$}

\

For this value of $\beta$, we can write $\alpha$ in terms of $\mu$ as $\alpha=-(1+\mu)$, while the mGCG fluctuations in Eq.~\rf{2.23} are
\ba{2.25}
	\delta\varepsilon_{\mathrm{Ch}}
		&=&
		\frac{1-A_s}{A_s}\frac{\left(1-A_s\right)^{-\frac{1}{\mu}}}{1+\mu}
		{
		\left(\frac{a_0}{a}\right)^{3-3\mu}
		\frac{\epschnow}{c^2 a}
		}\varphi
		-
		\frac{\left(1-A_s\right)^{-\frac{1}{\mu}}}{\mu(1+\mu)}
		{
		\left(\frac{a_0}{a}\right)^{3}
		\frac{\epschnow}{c^2 a}
		}\varphi
		\nn \\
		&{}&+
		\frac{1}{3(1+\mu)}\frac{1-A_s}{A_s}
		{
		\left(\frac{a_0}{a}\right)^{-3\mu}
		}\delta\varepsilon_{\mathrm{rad2}}
		+
		\frac{1}{3(1+\mu)}\delta\varepsilon_{\mathrm{rad2}}
		\, .
\ea
Then, Eq.~\rf{2.4} is consistent if the following system of equations is compatible
\ba{2.26}
		&&
		\frac{1}{1+\mu}\left[
		\frac{4+3\mu}{3}
		{
		\frac{\kappa c^2 a^3}{2}
		}
		\delta\varepsilon_{\mathrm{rad2}}
		-\frac{(1-A_s)^{-\frac{1}{\mu}}}{\mu}
		{
		\left(\frac{a_0}{a}\right)^3\frac{\kappa a^2\epschnow}{2}
		}
		\varphi
		\right]
		 =0
		\, ,
		 \\
		&&
		\frac{1}{1+\mu}\frac{1-A_s}{A_s}
		\left[
		\frac{1}{3}
		{
		\left(\frac{a_0}{a}\right)^{-3\mu}\frac{\kappa c^2a^3}{2}
		}
		\delta\varepsilon_{\mathrm{rad2}}
		+(1-A_s)^{-\frac{1}{\mu}}
		{
		\left(\frac{a_0}{a}\right)^{3-3\mu}\frac{\kappa a^2 \epschnow}{2}
		}\varphi
		\right]
		=0\label{2.27}
\, .
\ea
{ As we have already mentioned in subsection 3.2, the solution $\delta\varepsilon_{\mathrm{rad2}}=\varphi=0$ is a non-physical one and should be excluded. Then, from equations \rf{2.26} and \rf{2.27} we get respectively
\be{2.27a}
\delta\varepsilon_{\mathrm{rad2}}=\frac{3}{4+3\mu}\frac{(1-A_s)^{-1/\mu}}{\mu}\left(\frac{a_0}{a}\right)^3\frac{\varphi}{c^2a}\epschnow
\ee
and
\be{2.27b}
\delta\varepsilon_{\mathrm{rad2}}=-3(1-A_s)^{-1/\mu}\left(\frac{a_0}{a}\right)^3\frac{\varphi}{c^2a}\epschnow\, .
\ee
The compatibility of these two equations results in the constraint $\mu(4+3\mu)=-1$. Obviously, this constraint is not fulfilled for $\mu>0$.
Therefore, the case $\xi<0$, $\beta=0$ is {\bf forbidden} within the mechanical approach.}

\

{\em iii) $\beta<0,\, \beta \neq -1$}

\

In this case the Chaplygin gas fluctuations have the following structure
\be{2.28}
\delta\varepsilon_{\mathrm{Ch}} \approx
	-\frac{1+\beta}{\beta}(1-A_s)^{\frac{1}{1+\alpha}}
	\left[
		1
		+\sum_{n=1}^k N_n\left(\frac{a_0}{a}\right)^{3n\mu}
	\right]
	\left(\frac{a_0}{a}\right)^{3(1+\beta)}
	{
	\frac{\epschnow}{c^2 a}
	}
	\varphi
	-\frac{1}{3\beta}\delta\varepsilon_{\mathrm{rad2}}\, ,
\ee
where $k\equiv[-\beta/\mu]$ is the smallest integer value of the ratio $-\beta/\mu$ and
\be{2.29}
	N_1 = -\frac{A_s}{1-A_s}
	\left[
		\frac{1+\beta}{\mu}
		- \frac{1+\mu}{\beta}
	\right]\, .
\ee

First, we consider the condition $\delta\varepsilon_{\mathrm{rad2}}\neq 0$. Because $\beta<0$, the $\delta\varepsilon_{\mathrm{rad2}}$ terms in \rf{2.28} and in \rf{2.4} cannot cancel each other. Therefore, the first expression on the rhs of Eq.~\rf{2.28} should have the structure $const.\times (1/a^4) + o(1/a^4)$. A simple analysis shows that this does not take place.

If $\delta\varepsilon_{\mathrm{rad2}} = 0$, then, for consistency of Eq.~\rf{2.4}, the first expression on the rhs of Eq.~\rf{2.28} should have the structure $const.\times (1/a^3) + o(1/a^4)$. This takes place for $\beta=-1/3$ and $\mu > 1/3$ implying $\xi <-1/3$.

\subsection{The Little Sibling case: $\beta=-1$, $\alpha\neq -1$}
\label{LSBR}

\

As we mentioned above, the case $\beta=-1$ is a special one. We begin with Eq.~\rf{2.1a}
\be{new8}
	\bar p_{\mathrm{Ch}}
		= - \epsch
		- \frac{\widetilde A}{\bar\varepsilon^{\alpha}_{\mathrm{Ch}}}
		\, ,
\ee
where $\widetilde{A} \neq 0$. This EoS generalises our recent work \cite{Bouhmadi-Lopez:2014cca} (see also Ref.~\cite{Bouhmadi-Lopez:2014gza}) and it leads to what we named there as the Little Sibling of the Big Rip singularity.
Integrating the conservation equation for the mGCG we obtain the background energy density
\be{2.30}
	\epsch = \epschnow
	\left[
		1
		+ \ln\left(\frac{a}{a_0}\right)^{3(1+\alpha)\widetilde A_s}
	\right]^{\frac{1}{1+\alpha}}
	\, ,
\ee
where $\widetilde A_s=\widetilde A /\epschnow^{1+\alpha}$.
For $\alpha=0$, we recover the case we studied in Ref.~\cite{Bouhmadi-Lopez:2014cca}.
If we require that the energy density is well defined, the rhs of the last equation must be positive. This is ensured by setting $(1+\alpha)^{-1}=2n$, where $n$ is an integer, or, for a general $\alpha$, by requiring that the quantity inside the square brackets be positive \footnote{We will disregard the fine-tuned case $(1+\alpha)^{-1}=2n$.}.
If we define the quantity
\ba{new3.32}
	\tilde{a}_*\equiv a_0\exp\left[-\frac{1}{3(1+\alpha)\widetilde A_s}\right] \, ,
\ea
we can rewrite Eq.~\rf{2.30} as
\ba{3.33}
	\epsch = \epschnow
	\left[
		\ln\left(\frac{a}{\tilde{a}_*}\right)^{3(1+\alpha)\widetilde A_s}
	\right]^{\frac{1}{1+\alpha}}
	\, .
\ea
Therefore, the quantity $\tilde{a}_*$ defines the minimum or maximum (depending on the sign of $(1+\alpha)\widetilde A_s$) allowed value of the scale factor for which the energy density of the mGCG is well defined. The results obtained are summarized in Table~\ref{behav_beta-1}.
\begin{table}[t]
\centering
	\begin{tabu}{ C{.47\textwidth} : C{.47\textwidth} }
	\( (1+\alpha)\widetilde A_s<0 \)		& \( 0<(1+\alpha)\widetilde A_s\) \\ \hline
	\( 0<a<\tilde a_* \) 	& \( \tilde a_*<a<+\infty \) \\
	
	\(\alpha<-1 \Rightarrow \lim\limits_{a\to\tilde a_*}\epsch(a) = +\infty\)
		& \( \alpha<-1 \Rightarrow \lim\limits_{a\to\tilde a_*}\epsch(a) = +\infty\) \\
	\( -1 < \alpha \Rightarrow \lim\limits_{a\to\tilde a_*}\epsch(a) = 0\)
		& \( -1 < \alpha \Rightarrow \lim\limits_{a\to\tilde a_*}\epsch(a) = 0 \) \\ \hline
\end{tabu}
\caption{\label{behav_beta-1}Behaviour of the mGCG for $\beta=-1$.}
\end{table}
In the mechanical approach, we require that the background quantities be well defined for arbitrarily large values of the scale factor (see footnote~\ref{footnote1} above), therefore we must restrict our analysis to the case $(1+\alpha)\widetilde A_s>0$.

From Eqs.~\rf{new8} and \rf{2.30}, we can easily obtain,
\ba{2.31}
	&{}&\bar{\varepsilon}_{\mathrm{Ch}}+\bar{p}_{\mathrm{Ch}}
		=-\widetilde A_s\epschnow^{ } \left[
		1
		+ \ln\left(\frac{a}{a_0}\right)^{3(1+\alpha)\widetilde A_s}
	\right]^{-\frac{\alpha}{1+\alpha}}
	\, ,\\
\label{2.32}
	&{}&\delta p_{\mathrm{Ch}}
		= \left(
			-1
			+
			\frac{\alpha \widetilde A_s}{1
			+\ln\left(\frac{a}{a_0}\right)^{3(1+\alpha)\widetilde A_s}}
		\right) \delta\varepsilon_{\mathrm{Ch}}
		\, .
\ea
Then, substituting the previous expressions in Eq.~\rf{2.3}, we get
\be{2.33}
	\delta\varepsilon_{\mathrm{Ch}}=
	-\frac{
		\widetilde A_s
		\left[
			1
			+\ln\left(\frac{a}{a_0}\right)^{3(1+\alpha)\widetilde A_s}
		\right]^{\frac{1}{1+\alpha}}
	}
	{
		1
		-\alpha \widetilde A_s
		+\ln\left(\frac{a}{a_0}\right)^{3(1+\alpha)\widetilde A_s}
	}
	{
	\frac{\epschnow}{c^2 a}
	}
	\varphi
	+\frac{1}{3}
	\frac{
		\left[
			1
			+\ln\left(\frac{a}{a_0}\right)^{3(1+\alpha)\widetilde A_s}
		\right]
	}
	{
		1
		-\alpha \widetilde A_s
		+\ln\left(\frac{a}{a_0}\right)^{3(1+\alpha)\widetilde A_s}
	}
	\delta\varepsilon_{\mathrm{rad2}}\, ,
\ee
which for big values of the scale factor, $a$, the same quantity can be approximated as follows:
\be{2.34}
	\delta\varepsilon_{\mathrm{Ch}}\approx
	-\widetilde A_s
	\left[
		1
		+\ln\left(\frac{a}{a_0}\right)^{3(1+\alpha)\widetilde A_s}
	\right]^{-\frac{\alpha}{1+\alpha}}
	{
	\frac{\epschnow}{c^2 a}
	}\varphi
	+\frac{1}{3}\delta\varepsilon_{\mathrm{rad2}}\, .
\ee
The radiation terms in Eqs.~\rf{2.34} and in \rf{2.4} result in a non zero combination $(4/3)\delta\varepsilon_{\mathrm{rad2}}\sim 1/a^4$.
Obviously, such combination cannot cancel the first term in \rf{2.34} both for $\alpha \neq 0$ due to the logarithmic behaviour (as the logarithmic functions and the power law functions are linearly independent) and, in the case $\alpha =0$, because the first term behaves as $1/a$. If we put $\delta\varepsilon_{\mathrm{rad2}}\equiv 0$, Eq.~\rf{2.4} is still inconsistent due to the first term in the rhs of Eq.~\rf{2.34}. This conclusion follows from the simple observation that in this case the term $a^3 \delta\varepsilon_{\mathrm{Ch}}$ depends on time both for $\alpha \neq 0$ and for $\alpha =0$. Therefore, the case $\beta=-1,\, \alpha\neq -1$ is {\bf forbidden}.


\section{Conclusion and discussion}
\label{Conc}

\

In our paper, we have studied the Universe at its late stage of evolution and at scales much smaller than the cell of uniformity size which is approximately 190 Mpc \cite{EZcosm2}. At such scales, the Universe is highly inhomogeneous and the galaxies and groups of galaxies play the role of the inhomogeneities. We have considered the Universe filled with a modified generalized Chaplygin gas. Its main properties have been reviewed in section~\ref{mGCG}. The qualitative behaviour of the Universe as well as the asymptotic behavior of the mGCG are summarised in Tables~\ref{mGCG_restrictions}-\ref{behav_beta-1}. Additionally, we also took into account the presence of radiation in our (perturbative) analysis.

The main aim of this paper is to investigate the compatibility of the mGCG models with the theory of scalar perturbations within the mechanical approach \cite{EZcosm1,EZcosm2}. This approach is adequate for the late Universe and works well for big values of the scale factor. Therefore, we consider the equations of the scalar perturbations in this limit and found the sets of parameters which make these equations consistent. As a result, we select the sets of parameters of the mGCG models compatible with the mechanical approach. We summarise all acceptable cases in Table~\ref{results_table}. Depending on the sign of the parameter $\xi$, these cases can be split into two groups. Positive $\xi$ provides late time acceleration of the Universe and negative $\xi$ results in deceleration or zero acceleration of the late Universe. The former group is for us the most interesting case because the Chaplygin gas models have been introduced precisely to describe the late acceleration of our Universe. Here, we found three possible choices of the parameter $\xi$ which are compatible with the mechanical approach. The cases $\xi >1$ (except $\xi = 4/3$) and $\xi=4/3$ look very unnatural. In the first case the mGCG is not clustered in the presence of the inhomogeneities in the form of galaxies, even though so far we have no observational evidence that DE cluster. In the second case, the mGCG can be clustered. However, these fluctuations do not affect the gravitational potential because the rhs of \rf{2.11} is equal to zero. The third choice $\xi =2/3$ looks the most promising. Here, the mGCG is clustered and affect the gravitational potential.
\begin{table}[t]
\centering
\begin{tabu}{ C{.28\textwidth} : C{.20\textwidth} : C{.20\textwidth} : C{.20\textwidth} }
	 {\bf\large $\xi=(1+\alpha)(1+\beta) $}	
	 	& {\bf\large $\delta\varepsilon_{\mathrm{rad2}}$}
	 		& {\bf\large $\delta\varepsilon_{\mathrm{Ch}}$}
	 			&{\large late time} \\
\hline
	$\xi>1,\, \xi\neq 4/3$
		& $0$
			& $0$
				& cosm. constant\\
	$\xi=4/3$
		& can be non-zero
			& $\delta\varepsilon_{\mathrm{Ch}}\sim \delta\varepsilon_{\mathrm{rad2}}$
				& cosm. constant \\
	$\xi=2/3$
		& $0$
			& non-zero
				& cosm. constant\\
	$\xi<0,\, \beta>0,\, \beta\neq 1/3$
		& $0$
			& $0$
				& deceleration\\
	$\xi<0,\, \beta=1/3$
		& can be non-zero
			& $\delta\varepsilon_{\mathrm{Ch}}\sim \delta\varepsilon_{\mathrm{rad2}}$
				& deceleration\\
	$\xi<-1/3,\, \beta=-1/3$
		& $0$
			& non-zero
				& zero-acceleration\\
	\hline
\end{tabu}
\caption{\label{results_table}Values of the parameters of the mGCG model that are compatible with the mechanical approach.}
\end{table}

There are a number of papers where the latest observational data were used to get constraints on the mGCG models, considered there as a unified CDM and DE model. There is also a different approach where mGCG models are applied exclusively to describe DE. On the background level the difference between these models consists in the cosmological parameter $\Omega_m$. If the mGCG model describes both CDM and DE, then $\Omega_m \sim 0.046$ and this parameter corresponds to the baryonic matter. Such type of models are usually referred to as the Unified Dark Energy - Dark Matter scenarios.\footnote{\label{7}It is worth noting that the GCG models may suffer from the problem associated with unphysical oscillations or blow-up in the matter power spectrum \cite{Sandvik}. This problem in the GCG models can be avoided with the help of an unique decomposition into dark energy and dark matter \cite{Bento}. In addition, the growth of inhomogeneities in the CG and GCG models was investigated within the Zel'dovich approximation in the papers \cite{Bilic,Chimento}, while the growth of the matter perturbations in the mGCG model was studied at the linear level in \cite{Paul}. It was shown in these articles that the large-scale inhomogeneities evolve in good agreement with observations. It is also of interest to investigate a possibility of decomposing the mGCG into DE and DM in a similar fashion to the one carried out in \cite{Bento}.} If the mGCG model is responsible only for DE, then $\Omega_m\sim 0.3$ and it describes the baryonic and DM sectors. Here, non-interacting DM is inhomogeneously clustered according to the standard picture. In our paper, we did not use the parameter $\Omega_m$ for our analysis. Hence, the discrepancy between these two approaches does not affect the results in Table~\ref{results_table}. However, the experimental restrictions on the parameter $\xi$ can be different for different approaches. Therefore, we should keep in mind this possible discrepancy in $\xi$ based on the observational data obtained for different approaches.

The former approach where the mGCG models describe both CDM and DE was investigated, for example, in the papers \cite{obs1,obs2}. The authors of the paper \cite{obs1} have used the Union2 dataset of type supernovae Ia, the observational Hubble data, the cluster X-ray gas mass fraction, the baryon acoustic oscillation, and the cosmic microwave background data and found in the 1$\sigma$ estimation that $\xi \in [0.8492,1.4588]$. They considered a flat Universe model. More wide 1$\sigma$ estimation was obtained in \cite{obs2}: $\xi \in [0.795,2.032]$. Here, the authors have used the data for Type Ia supernovae, baryon acoustic oscillations and for the Hubble parameter. The Universe was not supposed {\em a priori} to be spatially flat. As we can see, the values of the parameter $\xi > 1$ belong to these intervals. Therefore, both cases with $\xi >1$ indicated in Table~\ref{results_table} are in agreement with these restrictions. However, the most reasonable value of $\xi$ from a physical point of view is $\xi =2/3 \approx 0.6667$, which is out of these intervals. Hence, we can conclude that the status/viability of the mGCG model within the mechanical approach is unclear. There are four possibilities. The first, and the most extremal one, is that the mGCG model is not viable for the description of DE and CDM. A second possibility is that the mGCG model is viable only for a rather limited period of time and cannot be applied in the future. That is, the mGCG equation of state \rf{2.1} is not a fundamental one. Third, the mGCG is not clustered, similarly to what happens with the cosmological constant. From a physical point of view, even though so far we have no observational evidence that DE clusters, this is not very natural. However, if it takes place, the mGCG model can be viable. There is also a fourth possibility. As we mentioned above, in Refs.~\cite{obs1,obs2}, the mGCG describes both DE and CDM while dust like matter is considered only in a baryonic form with the cosmological parameter $\Omega_m \sim 0.046$.
However, we can investigate a different scenario where the mGCG is related mainly or exclusively with DE. Here, the pressureless matter is a combination of baryonic matter and CDM and $\Omega_m \sim 0.3$.
 It is quite possible that such choice of $\Omega_m$ changes the 1$\sigma$ interval for $\xi$ in such a way that the physically reasonable value $2/3$ will belong to it. Then, such model can become viable. It is also worthwhile to investigate the effects that the latest data for the CMB has on the 1$\sigma$ interval for $\xi$ (similar to what was done in \cite{Bereiro} for the GCG model).
A future investigation will show which of these four above mentioned possibilities can be realized.


\section*{Acknowledgements}

\

The work of MBL is supported by the Portuguese Agency ``Funda\c{c}\~ao para a Ci\^encia e Tecnologia'' through an Investigador FCT Research contract, with reference IF/01442/2013/ CP1196/CT0001. She also wishes to acknowledge the support from the Portuguese Grants PTDC/FIS/111032/2009 and UID/MAT/00212/2013 and the partial support from the Basque government Grant No. IT592-13 (Spain) and the Spanish Grant No. FIS2014-57956-P.
J.M. is thankful to UPV/EHU for a PhD fellowship and UBI for the hospitality during the completion of part of this work and acknowledges the support from the Basque government Grant No. IT592-13 (Spain) and the Spanish Grant No. FIS2014-57956-P.
A.Zh. acknowledges the hospitality of CENTRA/IST and UBI during the completion of a part of this work. { We also want to thank Maxim Eingorn for his contribution during the initial phase of work.}



\begin{thebibliography}{99}
\bibitem{Planck2015}
P.~A.~R.~Ade {\it et al.} [Planck Collaboration],
 {\em Planck 2015 results. XIII. Cosmological parameters};
 \href{http://arxiv.org/abs/1502.01589}{arXiv:1502.01589 [astro-ph.CO]}.

\bibitem{Nojiri:2006ri}
 S.~Nojiri and S.~D.~Odintsov,
 {\em Introduction to modified gravity and gravitational alternative for dark energy},
eConf C {\bf 0602061} (2006) 06
 [Int.\ J.\ Geom.\ Meth.\ Mod.\ Phys.\ {\bf 4} (2007) 115];
 \href{http://arxiv.org/abs/hep-th/0601213}{arXiv:hep-th/0601213}.

\bibitem{DeFelice:2010aj}
 A.~De Felice and S.~Tsujikawa,
 {\em f(R) theories},
Living Rev.\ Rel.\ {\bf 13} (2010) 3;
 \href{http://arxiv.org/abs/1002.4928}{arXiv:1002.4928 [gr-qc]}.

\bibitem{Capozziello:2011et}
 S.~Capozziello and M.~De Laurentis,
 {\em Extended Theories of Gravity},
 Phys.\ Rept.\ {\bf 509} (2011) 167;
 \href{http://arxiv.org/abs/1108.6266}{arXiv:1108.6266 [gr-qc]}.

\bibitem[Clifton et al.(2012)]{2012PhR...513....1C}
T. Clifton, P.~G. Ferreira, A. Padilla, C. Skordis,
 {\em Modified gravity and cosmology},
 Phys.\ Rep.\ {\bf513} (2012) 1;
 \href{http://arxiv.org/abs/1106.2476}{arXiv:1106.2476 [astro-ph.CO]}

\bibitem{Morais:2015ooa}
 J.~Morais, M.~Bouhmadi-L\'opez and S.~Capozziello,
 {\em Can f(R) gravity contribute to (dark) radiation?},
 JCAP {\bf 1509} (2015) 09, 041;
 \href{http://arxiv.org/abs/1507.02623v2}{arXiv:1507.02623 [gr-qc]}.

\bibitem{Weinberg:1988cp}
 S.~Weinberg,
 {\em The Cosmological Constant Problem},
 Rev.\ Mod.\ Phys.\ {\bf 61} (1989) 1.

\bibitem{Peebles:2002gy}
 P.~J.~E.~Peebles and B.~Ratra,
 {\em The Cosmological constant and dark energy},
 Rev.\ Mod.\ Phys.\ {\bf 75} (2003) 559;
 \href{http://arxiv.org/abs/astro-ph/0207347}{arXiv:astro-ph/0207347}.

\bibitem{Padmanabhan:2002ji}
 T.~Padmanabhan,
 {\em Cosmological constant: The Weight of the vacuum},
 Phys.\ Rept.\ {\bf 380} (2003) 235;
 \href{http://arxiv.org/abs/hep-th/0212290}{arXiv:hep-th/0212290}.

\bibitem{Copeland:2006wr}
 E.~J.~Copeland, M.~Sami and S.~Tsujikawa,
 {\em Dynamics of dark energy},
 Int.\ J.\ Mod.\ Phys.\ D {\bf 15} (2006) 1753;
 \href{http://arxiv.org/abs/hep-th/0603057}{arXiv:hep-th/0603057}.

\bibitem{Frieman:2008sn}
 J.~Frieman, M.~Turner and D.~Huterer,
 {\em Dark Energy and the Accelerating Universe},
 Ann.\ Rev.\ Astron.\ Astrophys.\ {\bf 46} (2008) 385;
 \href{http://arxiv.org/abs/0803.0982}{arXiv:0803.0982 [astro-ph]}.

\bibitem{Kamenshchik:2001cp}
 A.~Y.~Kamenshchik, U.~Moschella and V.~Pasquier,
 {\em An Alternative to quintessence},
 Phys.\ Lett.\ B {\bf 511} (2001) 265;
 \href{http://arxiv.org/abs/gr-qc/0103004}{arXiv:gr-qc/0103004}.

\bibitem{Bilic:2001cg}
 N.~Bili{\'c}, G.~B.~Tupper and R.~D.~Viollier,
 {\em Unification of dark matter and dark energy: The Inhomogeneous Chaplygin gas},
 Phys.\ Lett.\ B {\bf 535} (2002) 17;
 \href{http://arxiv.org/abs/astro-ph/0111325}{arXiv:astro-ph/0111325}.

\bibitem{Bento:2002ps}
 M.~C.~Bento, O.~Bertolami and A.~A.~Sen,
 {\em Generalized Chaplygin gas, accelerated expansion and dark energy matter unification},
 Phys.\ Rev.\ D {\bf 66} (2002) 043507;
 \href{http://arxiv.org/abs/gr-qc/0202064}{arXiv:gr-qc/0202064}.

\bibitem{Wang:2013qy}
 Y.~Wang, D.~Wands, L.~Xu, J.~De-Santiago and A.~Hojjati,
 {\em Cosmological constraints on a decomposed Chaplygin gas},
 Phys.\ Rev.\ D {\bf 87} (2013) 8, 083503;
 \href{http://arxiv.org/abs/1301.5315}{arXiv:1301.5315 [astro-ph.CO]}.

\bibitem{Borges:2013bya}
 H.~A.~Borges, S.~Carneiro, J.~C.~Fabris and W.~Zimdahl,
 {\em Non-adiabatic Chaplygin gas},
 Phys.\ Lett.\ B {\bf 727} (2013) 37;
 \href{http://arxiv.org/abs/1306.0917v2}{arXiv:1306.0917 [astro-ph.CO]}.

\bibitem{Carneiro:2014jza}
 S.~Carneiro and C.~Pigozzo,
 {\em ``Observational tests of non-adiabatic Chaplygin gas},
 JCAP {\bf 1410} (2014) 060;
 \href{http://arxiv.org/abs/1407.7812v2}{arXiv:1407.7812 [astro-ph.CO]}.

\bibitem{BouhmadiLopez:2004me}
 M.~Bouhmadi-L\'opez and J.~A.~Jim\'enez Madrid,
 {\em Escaping the big rip?},
 JCAP {\bf 0505} (2005) 005;
 \href{http://arxiv.org/abs/astro-ph/0404540}{arXiv:astro-ph/0404540}.


 \bibitem{Bouhmadi;fraz1}
 M. Bouhmadi-L\'opez, P. Fraz\~ao, A. B. Henriques,
 {\em Stochastic gravitational waves from a new type of modified Chaplygin gas},
 Phys.\ Rev.\ D\ {\bf81} (2010) 063504;
 \href{http://arxiv.org/abs/0910.5134v2}{arXiv:0910.5134 [astro-ph.CO]};
 M. Bouhmadi-L\'opez, P. Fraz\~ao, A. B. Henriques,
 {\em Gravitons production in a new type of GCG};
 \href{http://arxiv.org/abs/1002.4785v1}{arXiv:1002.4785 [astro-ph.CO]}.

 \bibitem{Bouhmadi;chen}
 M. Bouhmadi-L\'opez, P. Chen, Y.-W. Liu,
 {\em Cosmological Imprints of a Generalized Chaplygin Gas Model for the Early Universe},
 Phys.\ Rev.\ D\ {\bf84} (2011) 023505;
 \href{http://arxiv.org/abs/1104.0676}{arXiv:1104.0676 [astro-ph.CO]}.

 \bibitem{Bouhmadi;mor1}
 M. Bouhmadi-L\'opez, J. Morais, A. B. Henriques, {\em Smoking guns of a bounce in modified theories of gravity through the spectrum of the gravitational waves},
 Phys.\ Rev.\ D\ {\bf87} (2013) 103528;
 \href{http://arxiv.org/abs/1210.1761v2}{arXiv:1210.1761 [astro-ph.CO]};
 M.~Bouhmadi-L\'opez, J.~Morais and A.~B.~Henriques,
 {\em The spectrum of gravitational waves in an f(R) model with a bounce}, Springer Proc.\ Math.\ Stat.\ {\bf 60} (2014) 157;
 \href{http://arxiv.org/abs/1302.2038v1}{arXiv:1302.2038 [astro-ph.CO]}.

 \bibitem{Bouhmadi;chen2}
 M. Bouhmadi-L\'opez, P. Chen, Y.-C. Huang, Y.-H. Lin,
 {\em Slow-Roll Inflation Preceded by a Topological Defect Phase \`a la Chaplygin Gas},
 Phys.\ Rev.\ D\ {\bf87} (2012) 103513;
 \href{http://arxiv.org/abs/1212.2641v2}{arXiv:1212.2641v2 [astro-ph.CO]}.


\bibitem{Benaoum:2002zs}
 H.~B.~Benaoum,
 {\em Accelerated universe from modified Chaplygin gas and tachyonic fluid};
 \href{http://arxiv.org/abs/hep-th/0205140}{arXiv:hep-th/0205140}.

\bibitem{Nojiri:2005sx}
 S.~Nojiri, S.~D.~Odintsov and S.~Tsujikawa,
 {\em Properties of singularities in (phantom) dark energy universe},
 Phys.\ Rev.\ D {\bf 71} (2005) 063004;
 \href{http://arxiv.org/abs/hep-th/0501025v2}{arXiv:hep-th/0501025}.

\bibitem{BouhmadiLopez:2006fu}
 M.~Bouhmadi-L\'opez, P.~F.~Gonz\'alez-D\'iaz and P.~Mart\'in-Moruno,
 {\em Worse than a big rip?},
 Phys.\ Lett.\ B {\bf 659} (2008) 1;
 \href{http://arxiv.org/abs/gr-qc/0612135v2}{arXiv:gr-qc/0612135}.

\bibitem{BouhmadiLopez:2007qb}
 M.~Bouhmadi-L\'opez, P.~F.~Gonz\'alez-D\'iaz and P.~Mart\'in-Moruno,
 {\em On the generalised Chaplygin gas: Worse than a big rip or quieter than a sudden singularity?},
 Int.\ J.\ Mod.\ Phys.\ D {\bf 17} (2008) 2269;
 \href{http://arxiv.org/abs/0707.2390v2}{arXiv:0707.2390 [gr-qc]}.

\bibitem{EZcosm1}
M. Eingorn and A. Zhuk,
 {\em Hubble flows and gravitational potentials in observable Universe},
 JCAP {\bf 09} (2012) 026;
 \href{http://arxiv.org/abs/1205.2384}{arXiv:1205.2384 [astro-ph.CO]}.

\bibitem{EKZ2}
M. Eingorn, A. Kudinova and A. Zhuk,
 {\em Dynamics of astrophysical objects against the cosmological background},
 JCAP {\bf 04} (2013) 010;
 \href{http://arxiv.org/abs/1211.4045}{arXiv:1211.4045 [astro-ph.CO]}.

\bibitem{EZcosm2}
M. Eingorn and A. Zhuk,
 {\em Remarks on mechanical approach to observable Universe},
 JCAP {\bf 05} (2014) 024;
 \href{http://arxiv.org/abs/1309.4924}{arXiv:1309.4924 [astro-ph.CO]}.

\bibitem{BUZ1}
A. Burgazli, M. Eingorn and A. Zhuk,
 {\em Rigorous theoretical constraint on constant negative EoS parameter $\omega$ and its effect for the late Universe\/},
 Eur.\ Phys.\ J.\ C {\bf 75} (2015) 3, 118;
 \href{http://arxiv.org/abs/1301.0418}{arXiv:1301.0418 [astro-ph.CO]}.

\bibitem{Laslo2}
M. Brilenkov, M. Eingorn, L. Jenkovszky and A. Zhuk,
 {\em Scalar perturbations in cosmological models with quark nuggets\/},
 Eur. Phys. J. C {\bf 74} (2014) 3011;
 \href{http://arxiv.org/abs/1310.4540}{arXiv:1310.4540 [astro-ph.CO]}.

\bibitem{ENZ1}
M. Eingorn, J. Nov$\mathrm{\acute{a}}$k and A. Zhuk,
 {\em $f(R)$ gravity: scalar perturbations in the late Universe\/},
 Eur. Phys. J. C {\bf 74} (2014) 3005;
 \href{http://arxiv.org/abs/1401.5410}{arXiv:1401.5410 [astro-ph.CO]}.

\bibitem{CPL}
\"{O}. Akarsu, M. Bouhmadi--L\'opez, M. Brilenkov, R. Brilenkov, M. Eingorn and A. Zhuk,
 {\em Are CPL models compatible with the history of our Universe?},
 JCAP {\bf 07} (2015) 038;
 \href{http://arxiv.org/abs/1502.04693v1}{arXiv:1502.04693 [gr-qc]}.

\bibitem{RGS}
{ We would like to thank Valery Rubakov, Dmitry Gorbunov and Sergey Sibiryakov who pointed to this important point.}

\bibitem{ChInfl}
B.R. Dinda, S. Kumar and A.A. Sen,
{\em GCG Inflation in the light of Planck and BICEP2},
Phys.\ Rev.\ D {\bf 90} (2014) 8, 083515;
\href{http://arxiv.org/abs/1404.3683}{arXiv:1404.3683 [astro-ph.CO]}.

\bibitem{Bouhmadi-Lopez:2014cca}
 M.~Bouhmadi-L\'opez, A.~Errahmani, P.~Mart\'in-Moruno, T.~Ouali and Y.~Tavakoli,
 {\em The little sibling of the big rip singularity},
 Int.\ J.\ Mod.\ Phys.\ D {\bf 24} (2015) 1550078;
 \href{http://arxiv.org/abs/1407.2446v1}{arXiv:1407.2446 [gr-qc]}.

\bibitem{Bouhmadi-Lopez:2014gza}
 M.~Bouhmadi-L\'opez, F.~S.~N.~Lobo and P.~Mart\'in-Moruno,
 {\em Wormholes minimally violating the null energy condition},
JCAP {\bf 11} (2014) 007;
 \href{http://arxiv.org/abs/1502.04693v1}{arXiv:1407.7758 [gr-qc]}.

\bibitem{Sandvik}
H. Sandvik, M. Tegmark, M. Zaldarriaga and I. Waga,
{\em The end of unified dark matter?},
Phys.\ Rev.\ D {\bf 69} (2004) 123524; 
\href{http://arxiv.org/abs/astro-ph/0212114v2}{arXiv:astro-ph/0212114}.

\bibitem{Bento}
M.C.Bento, O. Bertolami and A.A. Sen,
{\em The revival of the Unified Dark Energy-Dark Matter Model?}
Phys.\ Rev.\ D {\bf 70} (2004) 083519;
\href{http://arxiv.org/abs/astro-ph/0407239v3}{arXiv:astro-ph/0407239}.

\bibitem{Bilic}
 N.~Bili\'c, G.~B.~Tupper and R.~D.~Viollier,
{\em Unification of dark matter and dark energy: The Inhomogeneous Chaplygin gas},
 Phys.\ Lett.\ B {\bf 535} (2002) 17
 \href{http://arxiv.org/abs/astro-ph/0111325}{arXiv:astro-ph/0111325}.
 

\bibitem{Chimento}
L.P. Chimento and R. Lazkoz
{\em Large-scale inhomogeneities in modified Chaplygin gas cosmologies},
Phys.\ Lett.\ B {\bf 615} (2005) 146;
\href{http://arxiv.org/abs/astro-ph/0411068}{arXiv:astro-ph/0411068}.

\bibitem{Paul}
B.C. Paul and P. Thakur,
{\em Observational constraints on modified Chaplygin Gas from cosmic growth},
JCAP {\bf 11} (2013) 052;
\href{http://arxiv.org/abs/1306.4808v1}{arXiv:astro-ph/1306.4808}.

\bibitem{obs1}
J.~Lu, L.~Xu, Y.~Wu and M.~Liu,
{\em Combined constraints on modified Chaplygin gas model from cosmological observed data: Markov Chain Monte Carlo approach},
Gen.\ Rel.\ Grav.\ {\bf 43} (2011) 819;
 \href{http://arxiv.org./abs/1105.1870v1}{arXiv:1105.1870 [astro-ph.CO]}.

\bibitem{obs2}
G.~S.~ Sharov,
{\em Observational constrains on cosmological models with Chaplygin gas and quadratic equation of state},
\href{http://arxiv.org./abs/1506.05246v1}{arXiv:1506.05246 [gr-qc]}.

\bibitem{Bereiro}
T. Barreiro, O. Bertolami and P. Torres
{\em WMAP5 constraints on the unified model of dark energy and dark matter},
Phys.\ Rev.\ D {\bf 78} (2008) 043530;
\href{http://arxiv.org/abs/0805.0731v1}{arXiv:astro-ph/0805.0731}.


\end{thebibliography}
\end{document}